\documentclass[aps,prl,preprint,preprintnumbers]{revtex4}
\usepackage{graphicx}

\def\to{\rightarrow}
\def\Re{{\rm Re}}
\def\Im{{\rm Im}}
\def\lsim{\buildrel < \over {_\sim}}
     
\def\sh{\hat{s}}

\def\pt{p_{\rm T}}

\def\ptmiss{/{\hbox{\kern-6pt $\pt$}}}
\def\Br{{\rm Br}}

\def\MS{\rm\overline{MS}}

\def\ca{{\cal A}}
\def\Sublead{{\rm \scriptscriptstyle SL}}
\def\Lead{{\rm \scriptscriptstyle L}}
\def\spa#1.#2{\left\langle#1\,#2\right\rangle}
\def\spb#1.#2{\left[#1\,#2\right]}
\def\fig#1{fig.~{\ref{#1}}}
\def\Fig#1{Figure~{\ref{#1}}}
\def\eqn#1{eq.~(\ref{#1})}

\begin{document}

\preprint{hep-ph/0302233\\}
\preprint{SLAC--PUB--9654}

\title{Resonance-continuum interference\\ in the di-photon Higgs 
signal at the LHC}

\author{Lance Dixon and M. Stewart Siu}
\thanks{Research supported by the US Department of Energy under contract
DE-AC03-76SF00515.}
\affiliation{Stanford Linear Accelerator Center\\
Stanford University, Stanford, CA 94309, USA}

\date{\today}

\begin{abstract}
A low mass Standard Model Higgs boson should be visible at the Large 
Hadron Collider through its production via gluon-gluon fusion
and its decay to two photons.  We compute the interference of
this resonant process, $gg \to H \to \gamma\gamma$, with the continuum
QCD background, $gg \to \gamma\gamma$ induced by quark loops.
Helicity selection rules suppress the effect, which is dominantly due to
the imaginary part of the two-loop $gg \to \gamma\gamma$ 
scattering amplitude.  The interference is destructive, but only of order
5\% in the Standard Model, which is still below the 10--20\% present 
accuracy of the total cross section prediction.  We comment on the potential
size of such effects in other Higgs models.
\end{abstract}


\maketitle


The Higgs boson is the lone undetected elementary particle of the
Standard Model (SM), and the only scalar~\cite{Higgs}.  In the SM, it 
accounts for the masses of the $W$ and $Z$ bosons, quarks and 
charged leptons, and its properties are completely fixed by its mass.
Its detection, and measurement of its properties, are among the prime 
goals of the Fermilab Tevatron and the CERN Large Hadron Collider (LHC).

There is a good chance that the Higgs boson will be quite light.
Its mass in the SM is bounded from above by precision 
electroweak measurements, $m_H \lsim 196$--$230$ GeV at 
95\% CL~\cite{HiggsRadCorr}.  
The lightest Higgs boson in the Minimal Supersymmetric Standard Model 
(MSSM) must have a mass below about 135 GeV~\cite{SusyHiggs}.
These upper limits are not far above the lower bounds established 
by direct searches in the process $e^+e^- \to HZ$ at LEP2.  
The lower bound on the Higgs mass in the SM is 114.1~GeV;  
it drops to 91.0~GeV in the MSSM because the $HZZ$ 
coupling can be suppressed~\cite{LEP2SMandMSSMLimit}.

With sufficient integrated luminosity, Run II of the Tevatron may be able 
to discover a low mass Higgs; otherwise the task will fall to the LHC. 
For $m_H < 140$ GeV, the most important mode at the LHC involves Higgs 
production via gluon fusion, $gg\to H$~\cite{ggHGeorgi}, 
followed by the rare decay into two photons,
$H\to\gamma\gamma$~\cite{HggVertex,Higgsgammagamma}.  Although this
mode has a very large continuum $\gamma\gamma$
background~\cite{Higgsgammagamma,DW}, the narrow width of the Higgs
boson, combined with the 1\% mass resolution achievable in
the LHC detectors, allows the background to be measured experimentally
and subtracted from a putative signal peak~\cite{ATLASCMSetc}.

The branching ratio information provided by the $\gamma\gamma$ signal
is limited by the accuracy of the cross section for inclusive Higgs
production, $\sigma_H \equiv \sigma(pp \to HX)$, 
because only the product $\sigma_H \times \Br(H\to\gamma\gamma)$
is measured experimentally.
The next-to-leading order QCD corrections to $\sigma_H$ 
(dominated by gluon fusion) are very large~\cite{NLOHiggs}.  
Recently $\sigma_H$ was computed at 
next-to-next-to-leading order (NNLO)~\cite{NNLOHiggs}, in the heavy top 
quark limit --- which is an excellent approximation to the exact NLO cross 
section~\cite{NLOHiggs} for $m_H < 200$ GeV.  Threshold logarithms
have also been resummed at next-to-next-to-leading logarithmic 
accuracy~\cite{CdFG}. 
The residual theoretical uncertainties for $\sigma_H$, estimated by varying 
renormalization and factorization scales, are currently of 
order 10--20\%.  (The uncertainty in $\Br(H\to\gamma\gamma)$ 
is dominated by that in the $H \to b\bar{b}$ partial width,
and is smaller, of order 6\%~\cite{DSZ}.) 
In comparison, the anticipated experimental uncertainty in 
$\sigma_H \times \Br(H\to\gamma\gamma)$ with 100~fb$^{-1}$ 
of integrated luminosity per LHC detector is about 10\% for
115~GeV~$ < m_H < 145$~GeV~\cite{ZKNR}.

\begin{figure}
\includegraphics[width=10.65cm]{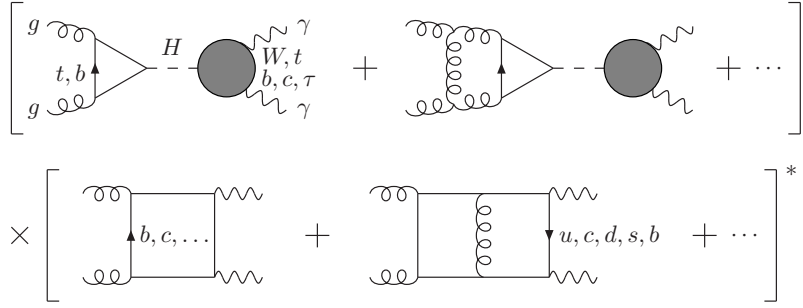}
\caption{\label{feyndiags} Sample Feynman diagrams contributing to 
the interference of $gg \to H \to \gamma\gamma$ with the continuum
background.  Only one diagram is shown at each loop order, for 
each amplitude.  The blob contains $W$ and $t$ loops, and small
contributions from lighter charged fermions.}
\end{figure}

It is critical to verify that no other physics alters the strength 
of the $\gamma\gamma$ signal at the 10\% level.  A potential
worry, addressed in this letter, is the interference between
the resonant Higgs amplitude $gg \to H \to \gamma\gamma$,
and the continuum $gg\to \gamma\gamma$ scattering process induced
by light quark loops.  Higgs resonance-continuum interference has been
studied previously in $gg\to H \to t\bar{t}$ at a hadron
collider~\cite{DSW}, and in $\gamma\gamma\to H \to W^+W^-$ and $ZZ$
at a photon collider~\cite{MTZNZK}.  These studies assumed that the 
Higgs boson is heavy enough to have a GeV-scale width.   In the case of 
a light ($m_H < 2 {\rm min}(m_W,m_t)$), narrow-width Higgs boson, the 
interference in $gg \to H \to \gamma\gamma$ was considered~\cite{DW}, 
but the dominant contribution in the SM was not identified.
Resonance-continuum interference effects are usually tiny for a narrow 
resonance, and for $m_H < 150$ GeV the width $\Gamma_H$ is less than 17~MeV.
However, the $gg \to H \to \gamma\gamma$ resonance is also rather weak.
As shown in \fig{feyndiags}, it consists of a one-loop production 
amplitude followed by a one-loop decay amplitude.  Thus a one-loop 
(or even two-loop) continuum amplitude can partially compete with it. 

In the SM, the production amplitude $gg \to H$ is dominated by a top quark 
in the loop.  The decay $H\to\gamma\gamma$ is dominated by the 
$W$ boson, with some $t$ quark contribution as well.
For $m_H < 160$~GeV, the Higgs is below the $t\bar{t}$ and $WW$ 
thresholds, so the resonant amplitude is mainly real, apart from the 
relativistic Breit-Wigner factor.  The full $gg\to\gamma\gamma$ amplitude 
is a sum of resonance and continuum terms,
\begin{equation}
\ca_{gg \to \gamma \gamma} = 
{ - \ca_{gg\to H} \ca_{H\to\gamma\gamma}
 \over \sh - m_H^2 + i m_H \Gamma_H } + \ca_{\rm cont} \,,
\label{rescontdecomp}
\end{equation}
where $\sh$ is the gluon-gluon invariant mass.  The interference term in
the partonic cross section is
\begin{eqnarray}
\delta\hat{\sigma}_{gg\to H\to \gamma\gamma} &=& 
-2 (\sh-m_H^2) { \Re \left( \ca_{gg\to H} \ca_{H\to\gamma\gamma} 
                          \ca_{\rm cont}^* \right) 
        \over (\sh - m_H^2)^2 + m_H^2 \Gamma_H^2 }
\nonumber\\
&& \hskip-0.3cm
-2 m_H \Gamma_H { \Im \left( \ca_{gg\to H} \ca_{H\to\gamma\gamma} 
                          \ca_{\rm cont}^* \right)
        \over (\sh - m_H^2)^2 + m_H^2 \Gamma_H^2 } \,.
\label{intpartonic}
\end{eqnarray}

At the hadron level, the interference term is
\begin{equation}
\delta\sigma_{pp\to H\to\gamma\gamma} = \int {d\sh \over \sh}
\, { dL_{gg} \over d\sh } 
\, \delta\hat{\sigma}_{gg\to H\to \gamma\gamma} \,,
\label{intcross}
\end{equation}
where the gluon-gluon luminosity function is 
\begin{equation}
{ dL_{gg} \over d\sh} 
 = \int_0^1 dx_1 dx_2 x_1 g(x_1) x_2 g(x_2) \delta(\sh/s - x_1 x_2). 
\label{ggLumi}
\end{equation}
The intrinsic Higgs width $\Gamma_H$ is much narrower than the 
experimental resolution $\delta m_H \sim 1$~GeV, so the observable
interference effect requires an integral across the entire linewidth.
The integral of the first, ``real'' term in \eqn{intpartonic} vanishes 
in the narrow-width approximation~\cite{DW} and leads to a subdominant
effect, to be discussed below.

The second, ``imaginary'', term in \eqn{intpartonic} has the same $\sh$
dependence as the resonance itself, so it survives integration over 
$\sh$ in the narrow-width limit (not counting the $\Gamma_H$ factor
already explicit in \eqn{intpartonic}).  However, it requires a
{\it relative phase} between the resonant and continuum amplitudes.
As mentioned above, in the SM the resonant amplitude, apart from the 
Breit-Wigner factor, is predominantly real.  The one-loop
continuum $gg\to\gamma\gamma$ amplitude is mediated by light quarks
in the loop.  Thus one might expect $\ca_{\rm cont}$ to have a 
large imaginary part, which is related by unitarity to the tree 
amplitude product 
$\ca_{gg\to q\bar{q}} \times \ca_{q\bar{q} \to \gamma\gamma}$.
For some gluon-photon helicity configurations this is true, but for 
the like-helicity cases $g^\pm g^\pm$ and $\gamma^\pm\gamma^\pm$
relevant for interference with a scalar Higgs resonance,
the tree amplitudes vanish as $m_q \to 0$~\cite{DW}.
At one loop, the imaginary part of $\ca_{\rm cont}$ comes
mainly from the $b$ and $c$ quark loops (as indicated in \fig{feyndiags}) 
and is suppressed by factors of order $e_q^2m_q^2/m_H^2$.

A much larger imaginary part of $\ca_{\rm cont}$ arises at the two-loop
order, where there is no quark mass suppression~\cite{GGGamGamtwoloop}.
In fact, the imaginary part of the two-loop $gg\to\gamma\gamma$ amplitude
is divergent due to an exchange of a soft-collinear virtual gluon between
the two incoming gluons, but this divergence cancels against a similar 
two-loop contribution to the production amplitude $\ca_{H\to gg}$.
We write the fractional interference correction to the resonance,
for polarized gluons and photons, as
\begin{eqnarray}
\delta &\equiv& { \delta\hat{\sigma} \over \hat\sigma } 
 = 2 m_H \Gamma_H\ \Im \Biggl[
{ \ca_{\rm cont}^{(1)} 
  \over \ca_{gg\to H}^{(1)} \ca_{H\to\gamma\gamma}^{(1)} }
\times \biggl( 1   
+ { \ca_{\rm cont}^{(2)} \over \ca_{\rm cont}^{(1)} }
- { \ca_{gg\to H}^{(2)} \over  \ca_{gg\to H}^{(1)} }
- { \ca_{H\to\gamma\gamma}^{(2)} \over  \ca_{H\to\gamma\gamma}^{(1)} }
\biggr) \Biggr] \,,
\label{Rinterf}
\end{eqnarray}
where for $\sh=m_H^2$~\cite{ggHGeorgi,HggVertex}
\begin{equation}
\ca_{gg\to H}^{(1)}
 = \sqrt{G_F\over 2\sqrt{2}} {\alpha_s(m_H) m_H^2 \over 3 \pi}
   \sum_{q=t,b,c} A_Q(4m_q^2/m_H^2)\,,
\label{ggHoneloop}
\end{equation}
\begin{eqnarray}
\ca_{H\to\gamma\gamma}^{(1)}
 &=& \sqrt{G_F\over 2\sqrt{2}} {\alpha m_H^2 \over 2 \pi}
  \biggl( 3 \sum_{q=t,b,c} e_q^2 A_{Q}^{\cal H}(4m_q^2/m_H^2)
\nonumber \\
&&
+ A_{Q}^{\cal H}(4m_\tau^2/m_H^2) + A_{W}^{\cal H}(4m_W^2/m_H^2)
  \biggr)\,,
\label{Hggoneloop}
\end{eqnarray}
with 
\begin{eqnarray}
 A_Q(x) &=& {3\over4} A_Q^{\cal H}(x) 
         = {3\over2} x ( 1 + (1-x) f(x) ) \,, 
\label{AQdef} \\
 A_W^{\cal H}(x) &=& - x \Bigl( 3 + {2\over x} + 3 (2-x) f(x) \Bigr) \,, 
\label{AWdef} \\
 f(x) &=& \cases{ (\sin^{-1}(1/\sqrt{x}))^2, & $x \geq 1$, \cr
            -{1\over4} \Bigl[ \ln\Bigl( 
    { 1+\sqrt{1-x} \over 1-\sqrt{1-x} } \Bigr) - i \pi \Bigr]^2 ,
    & $x < 1$. \cr}
\label{fdef}
\end{eqnarray}
Up to constant prefactors, the one-loop continuum amplitude 
$\ca_{\rm cont}^{(1)}$ for $gg\to\gamma\gamma$ is the same as for 
light-by-light scattering~\cite{GGGamGamoneloop,DW}, and is included 
with full quark mass dependence.  The two-loop amplitude 
$\ca_{\rm cont}^{(2)}$ is evaluated 
in the $m_q \to 0$ limit~\cite{GGGamGamtwoloop},
after cancelling the divergent terms in the ratio
$\ca_{\rm cont}^{(2)}/\ca_{\rm cont}^{(1)}$ against those in 
$\ca_{gg\to H}^{(2)}/\ca_{gg\to H}^{(1)}$.  The remaining
two-loop QCD corrections from $\ca_{gg\to H}^{(2)}$ and 
$\ca_{H\to\gamma\gamma}^{(2)}$ are included~\cite{SpiraReview}, but are 
small because they do not induce new phases.

A simplified approximate formula can be given by neglecting the 
remaining $\ca_{gg\to H}^{(2)}$ and $\ca_{H\to\gamma\gamma}^{(2)}$ terms, 
the small phase of $\ca_{\rm cont}^{(1)}$, and all but the 
(real) $W$ and $t$ loops in $\ca_{H\to\gamma\gamma}^{(1)}$ and 
$\ca_{gg\to H}^{(1)}$. 
There are two CP-inequivalent helicity configurations,
$g^+g^+ \to \gamma^+ \gamma^+$ and $g^-g^- \to \gamma^+ \gamma^+$.
However, the latter configuration continues to have vanishing imaginary
part at two loops, for massless quarks.  In terms of the functions 
$F^\Lead_{--++}$ and $F^\Sublead_{--++}$ used in 
ref.~\cite{GGGamGamtwoloop} to describe the former configuration, 
the correction in the unpolarized case is
\begin{eqnarray}
\delta &\approx& 
{ 2 \alpha \alpha_s^2(m_H) \, m_H \Gamma_H \sum_{q=u,c,d,s,b} e_q^2 
    \over \pi \, \Re( \ca_{gg\to H}^{(1)} ) 
                 \Re( \ca_{H\to\gamma\gamma}^{(1)} ) } 
\times
 \biggl( 3 \Im F^\Lead_{--++}(\theta) 
   - {1\over3} \Im F^\Sublead_{--++}(\theta) \biggr) \,,
\label{approxdelta}
\end{eqnarray}
where $\theta$ is the $gg \to \gamma\gamma$ center of mass scattering 
angle. 

\begin{figure*}
\includegraphics[width=11.90cm]{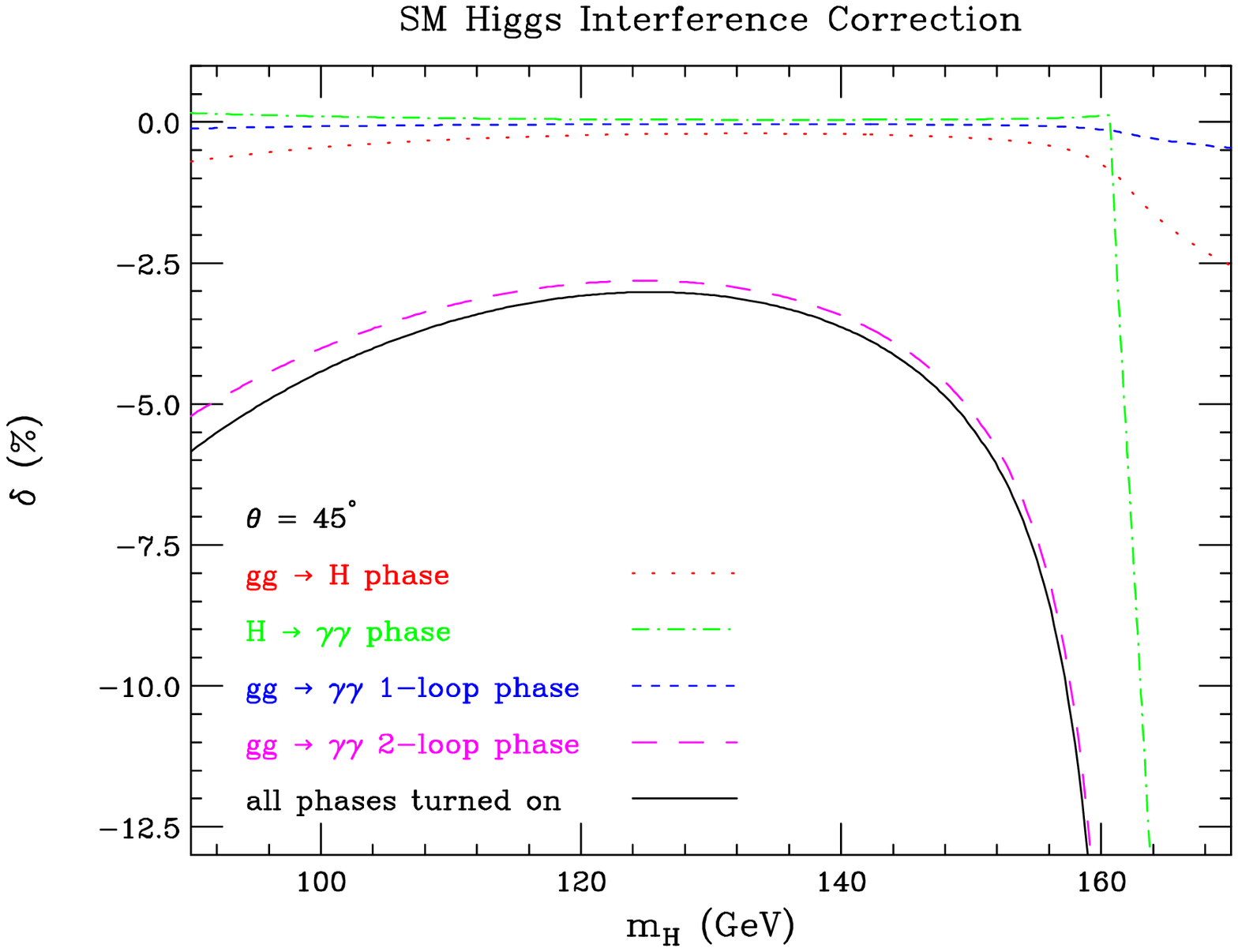}
\includegraphics[width=11.90cm]{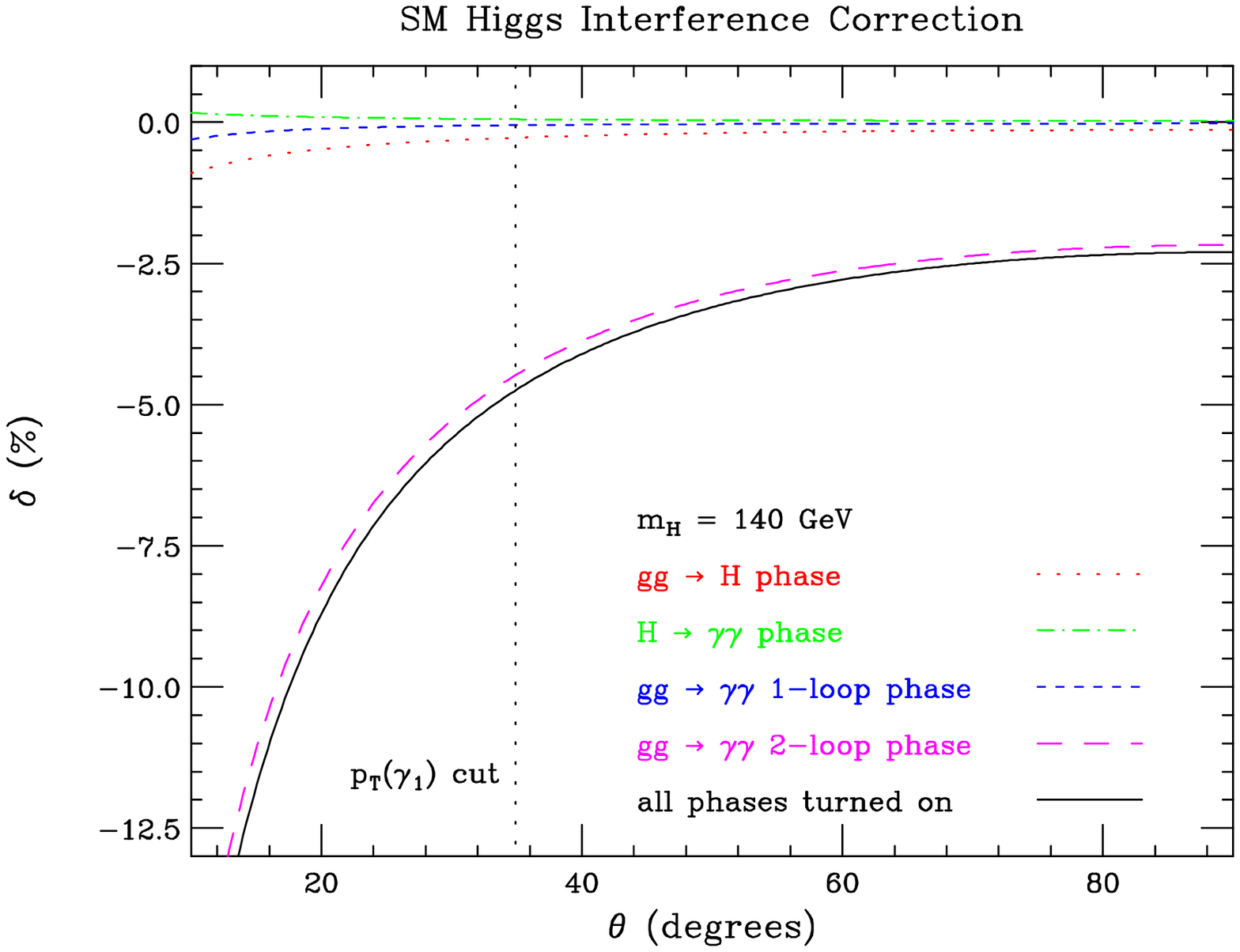}
\caption{\label{scanRavg} Top panel: the percentage reduction of the SM
Higgs $\gamma\gamma$ signal as a function of the Higgs mass, 
for CM scattering angle $\theta = 45^\circ$.  The solid curve gives 
the result with all phases turned on; the other curves turn on one of 
the component phases at a time.
Bottom panel:  the same quantities, plotted as a function of the
scattering angle, for $m_H = 140$~GeV.  The vertical dotted line indicates
that an event with $\theta < 34.9^\circ$ will not pass the standard 
ATLAS and CMS photon $\pt$ cuts.}
\end{figure*}

\Fig{scanRavg} shows the result of evaluating the unpolarized version of
\eqn{Rinterf}. We let $\alpha = 1/137.036$, $\alpha_s(m_Z) = 0.119$, 
and use $\MS$ quark masses evaluated at $\mu=m_H$, 
with $m_t(m_t) = 164.6$~GeV, $m_b(m_b) = 4.24$~GeV.  Our program
for Higgs boson decay widths is in good agreement with ref.~\cite{HDECAY}.
The left panel of \fig{scanRavg} plots $\delta$ as a function of $m_H$,
for $\theta = 45^\circ$.
The solid curve is the full result, while four other dashed and dotted
curves illustrate the result with one source of phase turned on at a time.
The effect is dominated by the phase arising from the imaginary part
of the two-loop continuum amplitude, for the helicity
configuration $g^+g^+ \to \gamma^+ \gamma^+$, as given 
by \eqn{approxdelta}. Not surprisingly,
it is smallest in the region the $\gamma\gamma$ signal is the strongest,
$100~{\rm GeV} < m_H < 140~{\rm GeV}$.  As $m_H$ increases toward $2m_W$,
the channel $H\to WW^*$ opens up, so $\Gamma_H$ and hence $\delta$ rise
rapidly.  The large phase arising from $\ca_{H\to\gamma\gamma}$ for 
$m_H > 2m_W$ is visible in the plot; however, such a signal will not 
be visible at the LHC.

The right panel of \fig{scanRavg} gives the $\theta$ dependence
of $\delta$, for $m_H = 140$~GeV.  The imaginary part of the 
continuum amplitude is forward peaked, so the effect rises 
there.  But the incoherent $q\bar{q}\to\gamma\gamma$ background
is also forward peaked, and so the experimental searches focus on
central scattering angles.   Indeed, at $m_H = 140$~GeV, an event
with $\theta < 34.9^\circ$ and no gluon radiation will produce
photons with transverse momentum $\pt(\gamma_{1,2}) < 40$~GeV,
below the standard ATLAS and CMS $\pt$ cuts~\cite{ATLASCMSetc}.

At the same order in $\alpha_s$ as the virtual corrections to 
$gg\to H\to\gamma\gamma$ represented by \eqn{Rinterf}, there are 
radiative corrections from the process $gg \to H \to \gamma\gamma g$
interfering with the one-loop $gg \to \gamma\gamma g$ continuum amplitude
induced by light quarks.
We evaluate the resonant amplitude in the heavy top approximation 
(see {\it e.g.} ref.~\cite{ggHg}), neglecting its small absorptive part,
and take the absorptive part of the
continuum amplitude for five massless quarks~\cite{gggamgamg}.
In the unpolarized cross section, only three CP conjugate pairs
contribute, due to helicity selection rules.  We convolute the 
interference term with standard gluon distributions, and integrate
over the final-state gluon momentum numerically,
with realistic rapidity and $\pt$ cuts on the photons.  The
result is remarkably miniscule compared to the virtual correction, 
amounting to 0.01\% or less of the signal.

Finally, we return to the ``real'' term in \eqn{intpartonic}.  It contains
the factor $\sh - m_H^2$ which is odd about $m_H$.  The resulting dip-peak
structure vanishes under integration~\cite{DW}, provided that the
nonresonant functions of $\sh$ vary slowly enough.  We perform a
first-order Taylor expansion of these functions about $m_H$, which
introduces a linear dependence on the cutoff (mass resolution) into the
integral.  For a resolution of 1 GeV, the integral of the real term in
\eqn{intpartonic} is negligible, representing 0.1\% or less of the
$\gamma\gamma$ signal over the region where it is visible.  The
contribution rises to a few percent for $m_H$ very near $2m_W$, where
$\ca_{H\to\gamma\gamma}$ has a sharp energy dependence (which is likely to
be smoothed out by finite $\Gamma_W$ effects).  At this large a Higgs
mass, however, the $H\to\gamma\gamma$ signal is unobservable.

Nonstandard Higgs sectors or other particle content could in 
principle generate a larger interference effect.  For example, the
Higgs coupling to the $b$ quark and $\tau$ lepton can be greatly 
enhanced in two-Higgs doublet models, including the MSSM.  This will 
increase the size of the phases of $\ca_{gg\to H}$ and
$\ca_{H\to\gamma\gamma}$ in \fig{scanRavg}.  However, these phases are
subdominant to the phase of $\ca_{\rm cont}^{(2)}$ in the SM, 
so the largest effect on \eqn{Rinterf} may come from an
increase in $\Gamma_H$ due to the $Hb\bar{b}$ coupling.
Yet if $\Gamma_H$ increases, the $H\to\gamma\gamma$ branching ratio
typically decreases, making this mode more difficult to detect and measure
accurately.  A more quantitative study is in progress~\cite{DSinProgress}.

Could other Higgs production and decay processes have appreciable
interference effects?  At hadron or lepton colliders, the process 
$gg\to H\to\gamma\gamma$ is almost unique in proceeding only at two loops.  
The only other potential signal of this type is $gg\to H\to Z\gamma$.  
The same helicity selection rules prohibit a one-loop continuum phase, 
but allow a two-loop one, so we expect to find an effect of similar 
magnitude, once the two-loop $gg\to Z\gamma$ amplitude is computed.  
The photon collider process $\gamma\gamma \to H \to \gamma\gamma$
will be discussed elsewhere; the corrections are below 
1\%~\cite{DSinProgress}.
Returning to the LHC, weak boson fusion followed by $H\to WW^*$ proceeds 
at tree level.  However, the $Z$ resonance can produce a significant 
phase in the one-loop continuum $W^*W^* \to WW^*$ amplitude, so this 
case may deserve investigation as well.

In summary, we have computed the dominant continuum interference 
corrections to the di-photon signal for the Standard Model Higgs boson
produced via gluon fusion.  The effects are at the 2--6\% level, depending
on the Higgs mass and scattering angles.  While still small compared to 
present theoretical and anticipated experimental errors, they are not
totally negligible, and suggest that further study is warranted of similar
effects in nonstandard models and for selected other channels.


\begin{acknowledgments}
We are grateful to Zvi Bern for suggesting this work.
We also thank Stefano Catani, Howard Haber, Maria Krawczyk, 
Zoltan Kunszt, Stefano Moretti, Michael Pes\-kin and Dieter Zeppenfeld
for helpful discussions.
\end{acknowledgments}


\end{document}